\documentclass[
 reprint,
 amsmath, amssymb,
 aps,prd
]{revtex4-2}

\usepackage{graphicx}%
\usepackage{dcolumn}%
\usepackage{bm}%
\usepackage{cancel} %
\usepackage[dvipsnames]{xcolor} %

\usepackage{xurl}

\usepackage[colorlinks=true,
            linkcolor=MidnightBlue,
            citecolor=ForestGreen,
            urlcolor=RoyalBlue]{hyperref}

\usepackage[nameinlink,noabbrev]{cleveref} 

\begin{document}

\preprint{APS/123-QED}

\def\apj{Astrophys.\ J.}
\def\apjl{Astrophys.\ J.\ Lett.}
\def\araa{ARA\&A}
\def\mnras{Mon.\ Not.\ R.\ Astron.\ Soc.}
\def\mnrasl{Mon.\ Not.\ R.\ Astron.\ Soc.\ Lett.}
\def\aap{Astron.\ Astrophys.}
\def\lrr{Living\ Rev.\ Relativ.}
\def\JCAP{J.\ Cosmol.\ Astropart.\ Phys.}
\def\jcap{J.\ Cosmol.\ Astropart.\ Phys.}
\def\jhep{J.\ High\ Energ.\ Phys.}
\def\cqg{Class.\ Quantum\ Grav.}
\def\prd{Phys.\ Rev.\ D\;}
\def\prc{Phys.\ Rev.\ C\;}
\def\pr{Phys.\ Rep.\ C\;}
\def\epjc{Eur.\ Phys.\ J.\ C\;}

\newcommand{\cg}[1]{{\color{black}#1}}
\newcommand{\cgg}[1]{{\color{black}#1}}
\newcommand{\st}[1]{{\color{black}#1}}

\title{
Self-Interacting Dark-Matter Spikes and the Final-Parsec Problem: Bayesian constraints from the NANOGrav 15-Year Gravitational-Wave Background
}%

\author{Shreyas Tiruvaskar}
 \email{sti50@uclive.ac.nz}
\author{Chris Gordon}%
 \email{chris.gordon@canterbury.ac.nz}
\affiliation{%
 School of Physical and Chemical Sciences, University of Canterbury, Christchurch, New Zealand
}%


\begin{abstract}

 A self-interacting dark-matter (SIDM) density spike around merging supermassive black holes (SMBHs) may be able to supply the dynamical friction needed to shrink binaries from $\sim 1\, \mathrm{pc}$ to $\sim 10^{-2} \,\mathrm{pc}$, thereby resolving the long-standing ``final-parsec problem''. Embedding the binary-halo system in a cosmological population model, we evolve the inspiral under the combined influence of gravitational-wave (GW) emission and SIDM drag, compute the resulting nanohertz GW background, and confront it with the NANOGrav 15-year pulsar-timing data. A six-parameter Bayesian analysis, performed with a Gaussian-process-accelerated Markov chain Monte Carlo, yields posterior constraints on the cross-section per unit mass and maximum circular velocity values that were consistent with independent galaxy-rotation and cluster-lensing limits. Within this parameter space, the SIDM spike remains intact, supplies sufficient friction to overcome the stellar depletion barrier, and produces a characteristic-strain spectrum that matches the NANOGrav signal as well as phenomenological astrophysical models.  
\end{abstract}

\maketitle

\section{\label{sec:intro}Introduction}

The “final‑parsec problem’’ arises because, in the simplest picture where gravitational‑wave (GW) emission is the only sink of orbital energy, a supermassive black‑hole binary (SMBHB) that has shrunk to a separation of $\sim1\,$pc still requires a merger time longer than a Hubble time~\cite{slingshot2}.  
At larger separations ($\gtrsim$ kpc) ordinary dynamical friction against stars and gas extracts energy efficiently, and three‑body slingshot interactions further harden the binary down to $\sim10$ pc~\cite{slingshot2,three_body_scattering_Sesana_2008,Kelley_2016}.  
Below this scale, however, stellar reservoirs are depleted and the inspiral stalls.  

Several purely baryonic remedies have been proposed. For example, triaxial or axisymmetric stellar potentials that continually repopulate loss cones~\cite{Khan_2013, Vasiliev_2014}, or torques from circum‑binary gas disks~\cite{Kocsis:2010xa,Goicovic:2016dul,Goicovic:2018xxi}. However, detailed simulations show that, in most galaxies, these mechanisms still leave coalescence times of several Gyr~\cite{Kelley_2016}.

 Using N-body simulations, Navarro, Frenk, and White  \cite{nfw_1996} found a spherically symmetric density profile for dark matter (DM) halos, which became known as the Navarro-Frenk-White (NFW) profile. However, it was proposed that dark matter might have a density profile, even steeper than the NFW one, due to its adiabatic accretion onto the central black hole \cite{silk_spike_1999}. This sharp rise in the DM density near the centre is known as the density ``spike''. 

\cg{These density spikes can provide enough dynamical friction \cite{chandrasekhar_1943} to dampen the inspiraling black hole binary and make them merge faster.}   However, if the dark matter in the spike is non-interacting, the kinetic energy that is transferred from the binary to the spike can lead to the ejection of the CDM particles, resulting in a disruption of the spike  \cite{alonso} (hereafter ACD2024). This could stall the binary, and the ``final parsec problem" would remain unsolved.
However, self-interactions in DM particles would lead to the redistribution of the transferred kinetic energy among the particles, which could prevent the disruption of the spike, providing a solution to the ``final parsec problem" (ACD2024).

Self-interacting dark matter (SIDM) was originally introduced to address some issues that cold dark matter (CDM) halo models face \cite{sidm_Spergel_2000}. One of the problems with the NFW profile for CDM is that it suggests a ``cuspy" (steep, centrally concentrated) profile, but observations of dwarf galaxies suggest flat-density cores \cite{core_cusp_Blok_2010, core_cusp_Governato_2010}. Interestingly, if self-interactions between the DM particles are considered, a flat density region is produced (known as the isothermal core) that matches the observations \cite{sidm_tulin_2018}. These observations can be used to constrain the key SIDM parameters, particularly the self-interaction cross-section (\(\sigma\)) and the SIDM particle mass (\(m\)). 

Observations of dwarf galaxies, galaxy clusters, and low surface brightness galaxies, when combined with numerical simulations of SIDM halos, provide constraints on the ratio \(\sigma/m\) \cite{sidm_dave_2000, sidm_elbert_2015, sidm_harvey_2019, sidm_review_adhikari2022, roberts2025}. Additionally, constraints can be derived from gravitational lensing measurements \cite{sidm_sagunski_2021} and X-ray observations of galaxy groups \cite{sidm_gopika_2023}. With recent observations of the gravitational wave background (GWB) data from pulsar timing array (PTA) experiments, a new avenue has emerged to probe the SIDM model through its signature on the GWB.

PTA experiments are sensitive to the GWB in the nanohertz frequency band \cite{gwb_nanograv, gwb_ppta, gwb_epta_2023, gwb_cpta_2023}. A key source of GWB in this regime is believed to be the merger of SMBHBs, which emit gravitational waves in the same frequency range \cite{gwb_smbh_Begelman_1980, Agazie_2023, alonso}. Therefore, any modifications to the SMBHB merger process may leave observable imprints in the PTA data. 
Several such modifications have been explored: orbital eccentricity alone \cite{Bi:2023tib,chen_2024}, generic environmental hardening \cite{Agazie_2023,Ellis:2023dgf}, cored or cusped DM halos \cite{Ghoshal:2023fhh}, dense DM spikes \cite{Shen:2023pan,Hu:2023oiu}, and \cg{ultra-light-DM  \cite{PhysRevD.105.083008,Aghaie:2023lan,2024PhRvD.110b3517B,2024PhRvD.110b3517B,2024JCAP...06..024B,PhysRevD.108.103517,2024PhRvD.110h4012M,2024PhRvL.132u1401A,Sarkar:2025tiy,Ding:2025nxe}}.
If an SIDM halo surrounds the SMBHB, it would provide dynamical friction, damping the motion of the SMBHB and altering the resulting gravitational wave (GW) signal (ACD2024). Thus, the effects of the SIDM halo on the SMBHB merger can be tested using the observed PTA data. 

In this work, we use recent GWB observations from NANOGrav \cite{gwb_nanograv} to place constraints on SIDM parameters. To achieve this, we begin by constructing a density profile for SIDM halos around SMBHs, which is then used to calculate the dynamical frictional force provided by the SIDM. Thus, we can obtain the energy and the power lost by the binary due to dynamical friction. Inspiraling SMBHB emits GWs, which leads to additional power loss. With the total power, we can calculate the decay rate of the binary separation. This leads us to numerically compute the characteristic strain spectrum and fit it to the strain spectra from the NANOGRav 15-year dataset \cite{gwb_nanograv}. For the statistical analysis, we use the Bayesian method of Markov chain Monte Carlo (MCMC), which employs simulated strain spectra and the NANOGrav data to give a posterior distribution of the model parameters.

The rest of the paper is structured as follows. In Section \ref{sec:density}, we give details of computing the SIDM density profile and how self-interactions affect it. Section \ref{sec:dynamics} describes the dynamics of the SMBHB merger and the effects of dynamical friction on the merger. In Section \ref{sec:strain}, we discuss how to calculate the characteristic strain numerically. Section \ref{sec:stats} outlines the statistical framework we use for the Bayesian analysis, i.e., the process of generating a library, training a Gaussian process, and using that to produce the corresponding MCMC. In Section \ref{sec:results}, we present the results of our work, with a focus on the constrained SIDM model parameters. Finally, we conclude with Section \ref{sec:conclusion} to discuss the other relevant contemporary studies and potential directions for future work.

\section{\label{sec:density}Density profile of a self-interacting dark matter halo}
In our model, the spherically symmetric NFW profile is modified near the centre to become a density spike, resulting from the accretion of SIDM onto the SMBH. Additionally, self-interactions between SIDM particles lead to the development of a flat isothermal core in the intermediate region. This leads to a spherically symmetric density profile that has three distinct regions: a central density spike, a flat-density isothermal core in the intermediate region, and an outer region that follows the NFW profile.

\subsection{NFW parameters}
The NFW density profile can be written using its two parameters as
\begin{equation}
    \rho(r) = \frac{\rho_s}{\left(r / r_{s}\right)\left(1+r / r_{s}\right)^{2}},
\label{nfw_profile}
\end{equation}
where \(r_s\) is the scaling radius and \(\rho_s\) is the NFW density parameter. These parameters can be found following the method in \cite{nfw1997}. Eq.~(1) in \cite{nfw1997} gives
\begin{equation}
    \frac{\rho(r)}{\rho_{\text {crit }}}=\frac{\delta_{c}}{\left(r / r_{s}\right)\left(1+r / r_{s}\right)^{2}},
\label{nfw1997_eq1}
\end{equation}
where \(\rho_{\text {crit }}\) is the critical density given by
\begin{equation}
    \rho_{\text {crit }}= \frac{3 H^{2}}{8 \pi G}.
\end{equation}
The critical density depends on redshift \(z\) via \(H\) as 
\begin{equation}
    H(z) = H_0 \sqrt{\Omega_m (1+z)^3 + \Omega_{\Lambda}},
\label{Hz}
\end{equation}
where  \(H_0\), \(\Omega_m\), and \(\Omega_{\Lambda}\) are the cosmological parameters. We follow the NANOGrav  \cite{Agazie_2023} (hereafter Agazie2023) approach and use the WMAP9 values for \(H_0\), \(\Omega_m\), and \(\Omega_{\Lambda}\). We use \(H_0=69.33\,\mathrm{km\, s^{-1}\, Mpc^{-1}}\), \(\Omega_m = 0.288\) and \(\Omega_{\Lambda} = 0.712\) (see the nine-year data for WMAP+BAO+\(\mathrm{H_0}\) in Table 3 of \cite{wmap9}). Next, the quantity \(\delta_{c}\) is given by Eq.~(2) in \cite{nfw1997} as
\begin{equation}
    \delta_c=\frac{200}{3} \frac{C^3}{[\ln (1+C)-C /(1+C)]},
\end{equation}
where \({C}\) is the concentration parameter, and is related to \(r_s\) as \(C=r_{200} / r_{s}\). 
The concentration parameter can be calculated as a function of halo mass \(M_{200}\) and redshift using Eq.~(24) in \cite{klypin2016} as
\begin{equation}
    C(M_{200}, z) = C_0 \left( \frac{M_{200}}{10^{12}\, h^{-1} \mathrm{M}_\odot} \right)^{-\gamma} \left[ 1 + \left( \frac{M_{200}}{M_0} \right)^{0.4} \right].
\end{equation}
\(C_0, M_0\), and \(\gamma\) can be interpolated to a required \(z\) value using Table 2 in \cite{klypin2016}. Thus, using black hole mass \(M_{\mathrm{BH}}\) and redshift \(z\), we can find halo mass \(M_{200}\), allowing us to calculate the concentration parameter \(C\) and \(\delta_c\). Comparing Eq.~\eqref{nfw_profile} and Eq.~\eqref{nfw1997_eq1}, we can write
\begin{equation}
    \rho_s = \rho_{\mathrm{crit}}(z) \cdot \delta_c(M_{200}, z).
\end{equation}
For calculating \(r_s\), we use  
\begin{equation}
    M_{200} = 200 \cdot \frac{4}{3} \pi r_{200}^3 \cdot \rho_{\mathrm{crit}}(z).
\end{equation}
However, \(r_{200} = C \cdot r_{s}\), so substituting that gives us
\begin{equation}
    M_{200} = 200 \cdot \frac{4}{3} \pi \cdot \left(C(M_{200}, z) \cdot r_{s}\right)^3 \cdot \rho_{\mathrm{crit}}(z).
\end{equation}
Therefore,
\begin{equation}
    \begin{aligned}
        r_{s} &= \left(\frac{M_{200}}{200 \cdot \frac{4}{3} \pi \cdot C(M_{200}, z) ^3 \cdot \rho_{\mathrm{crit}}(z)}\right)^\frac{1}{3}.
    \end{aligned}
\end{equation}
Thus, if we know \(M_{200}\) and \(z\), we can calculate the NFW parameters \(\rho_s\) and \(r_s\).

\subsection{Halo mass and stellar mass from the black hole mass}
To find a relation between the black hole mass, the galaxy stellar mass, and the halo mass, we follow  Eq.~(21) of  Agazie2023.
\begin{equation}
    \log_{10}(M_{\mathrm{BH}} / \mathrm{M}_\odot) = \mu + \alpha_\mu \log_{10} \left( \frac{M_{\mathrm{bulge}}}{10^{11}\, \mathrm{M}_\odot} \right) + \mathcal{N}(0, \epsilon_\mu).
\label{mbh_mbulge}
\end{equation}
Here, we set \(\alpha_{\mu}\) to 1.1 following the fiducial value given in Table 2 of  Agazie2023. \(\mathcal{N}(0, \epsilon_\mu)\) is the Gaussian random scatter with zero mean and standard deviation of \(\epsilon_{\mu}\) in dex. \(M_{\mathrm{bulge}}\) is the fraction of the galaxy's stellar mass present in the stellar bulge, which is related to the stellar mass of the galaxy \(M_\star\) as
\begin{equation}
    M_{\mathrm{bulge}} = 0.615 \cdot M_{\star}.
\label{mbulge_mstar}
\end{equation}
The stellar mass relates to the halo mass as
\begin{equation}
    \frac{M_{\star}}{M_{200}}(z) = 2 A(z) \left[ \left( \frac{M_{200}}{M_A(z)} \right)^{-\beta(z)} + \left( \frac{M_{200}}{M_A(z)} \right)^{\gamma(z)} \right]^{-1},
    \label{eq:stellar_mass_relates_to_halo_mass}
\end{equation}
following Eq.~(6) of \cite{girelli2020}. In our analysis we use Eq.~(7)-(10) and values from Table 3 of \cite{girelli2020} to calculate \(A(z)\), \(M_A(z)\), \(\beta(z)\), and \(\gamma(z)\). Thus, we have
\begin{equation}
    M_{\star} = 2 A(z) \frac{M_{200}}{\left( \frac{M_{200}}{M_A(z)} \right)^{-\beta(z)} + \left( \frac{M_{200}}{M_A(z)} \right)^{\gamma(z)}}.
\label{mstar_mhalo}
\end{equation}
Which means that if we have halo mass \(M_{200}\), then using Eq.~\eqref{mstar_mhalo}, Eq.~\eqref{mbulge_mstar}, and Eq.~\eqref{mbh_mbulge} we can obtain \(M_{\mathrm{BH}}\). We can numerically invert these to obtain \(M_{200}\) from \(M_{\mathrm{BH}}\).

If we know the black hole mass (\(M_{\mathrm{BH}}\)) and redshift (\(z\)), we can calculate the halo mass (\(M_{200}\)). Then, using the halo mass, we can calculate the NFW parameters (\(\rho_s\) and \(r_s\)).

\subsection{\label{subsec:core}SIDM core}
In the SIDM density profile, going radially outwards, the part that comes before the NFW profile is the isothermal core. Self-interactions allow the dark matter to form an isothermal core where the density varies very slowly with distance. In the core, density follows the Poisson equation (Eq.~(B1) in  ACD2024)
\begin{equation}
    v_0^2 \nabla^2 \ln \rho = -4 \pi G \rho,
\label{poisson}
\end{equation}
where \(v_0\) is the velocity dispersion of the SIDM particles, which is constant in the isothermal core. In a spherically symmetric DM halo, the \(v_0\) can also be thought of as the RMS speed of the DM particles. Boundary conditions for this differential equation are \(\rho'(0) = 0\) and \(\rho_c = \rho_{\mathrm{NFW}}(r_1)\), which are used to solve the Poisson equation and obtain \(\rho(r)\). Here, \(r_1\) is the distance from the centre where the isothermal core ends and the NFW profile begins. Note that a prime indicates a derivative with respect to the radial coordinate, i.e.\ $\rho'(r)\equiv\mathrm{d}\rho/\mathrm{d}r$.

 To determine \(r_1\), we use a condition that during the lifetime of the core (\(t_{\mathrm{age}}\)), there should be at least one scattering per SIDM particle within the core. i.e. for \(r<r_1\),
\begin{equation}
    \frac{\langle \sigma v \rangle}{m} \cdot \rho_{\mathrm{NFW}}(r_1) \cdot t_{\mathrm{age}} \sim 1.
\label{core_radius}
\end{equation}
Here, \(\sigma\) is the self-interaction cross section and \(m\) is the mass of the DM particle. The term \(\langle \sigma v \rangle\) is explained in section \ref{self-interactions}.

Moreover, we also apply the mass constraint condition on the density profile \(\rho(r)\) that the mass enclosed within \(r_1\) for the core profile should be equal to the mass that would have been enclosed within \(r_1\) by just the NFW profile.

Thus, in the calculation of isothermal core profile, if we know \((\sigma/m) \cdot t_{\mathrm{age}}\), we have two unknowns, the velocity dispersion \(v_0\) and the core radius \(r_1\). By simultaneously solving the mass constraint and Eq.~\eqref{core_radius} using numerical methods, we can find \(v_0\) and \(r_1\). Using that in Eq.~\eqref{poisson}, we can solve the Poisson equation utilising the boundary conditions, which gives us the core density profile \(\rho(r)\).

\subsection{SIDM spike}
The innermost part of the SIDM density profile is the DM spike, where the density rises rapidly due to accretion. The equation for the spike density profile is given as Eq.~(1) in  ACD2024 as
\begin{equation}
    \rho_{\mathrm{sp}}(r) = \rho_{\mathrm{sp}0} \left( \frac{r_{\mathrm{sp}}}{r} \right)^\gamma,
\label{spike}
\end{equation}
where \(\rho_{\mathrm{sp}0}\) is the constant scaling density and \(r_{\mathrm{sp}}\) is the spike radius. We follow the procedure in  ACD2024 and set \(\rho_{\mathrm{sp}0}=\) \(\rho_{\mathrm{core}}(r=0)\). In modelling the SIDM spike, ACD2024 uses results from ref.~\cite{ShapPas2014}. The spike radius is given by
\begin{equation}
    r_{\mathrm{sp}} = \frac{G M_{\mathrm{BH}}}{v_0^2}.
\label{spike_radius}
\end{equation}
Thus, if we know the black hole mass and the redshift, we can calculate NFW parameters, then the isothermal core parameters, and using that, the spike parameters \(\rho_{\mathrm{sp}0}\) and \(r_{\mathrm{sp}}\).

The remaining spike parameter, \(\gamma\), depends on what kind of self-interaction is considered between the SIDM particles.

\subsection{Self-interactions}
\label{self-interactions}
There can be different types of self-interactions between the SIDM particles, similar to baryonic matter. It could be a simple contact interaction (isotropic scattering). In that case, the cross-section does not depend on the velocity. However, if we consider an interaction like Coulomb scattering (mediated by a massless force carrier), there is a \(v^{-4}\) velocity dependence. This velocity dependence in these two cases can be written as
\begin{equation}
    \sigma \propto \frac{1}{v^a},
\label{sigma_v_dep}
\end{equation}
where \(a=0\) for the contact interaction, and \(a=4\) for the Coulomb scattering. Neglecting baryonic/stellar effects, the spike density exponent \(\gamma\) from Eq.~\eqref{spike} depends on the type of interaction as (ACD2024)
\begin{equation}
    \gamma = \frac{3 + a}{4}.
\label{gamma}
\end{equation}
Thus, for contact interaction, \(\gamma=3/4\) and for the Coulomb interaction, \(\gamma=7/4\). This means that for the Coulomb interaction (\(a=4\)), the spike is steeper than the contact interaction (\(a=0\)).

We consider an interaction where the mediator is a massive force carrier. Following  ACD2024, in the case of the massive mediator, the interaction type can change from \(a=0\) to \(a=4\) when the SIDM velocity becomes equal to the transition velocity \(v_t\). If the DM velocity is smaller than the transition velocity (\(v<v_t\)), the type of interaction is \(a=0\). However, if the DM velocity is larger than the transition velocity (\(v>v_t\)), the interaction is \(a=4\). The transition velocity is the free parameter of the model that we vary and probe in our analysis.

In the isothermal core, the DM velocity (the RMS velocity) is constant and denoted by \(v_0\). However, in the spike region (the innermost region), the velocity increases as we move closer to the centre. We follow Eq.~(B6) in  ACD2024, that states
\begin{equation}
    \frac{v(r)}{v_0} = \frac{7}{11} + \frac{4}{11} \left( \frac{r_{\text{sp}}}{r} \right)^{1/2}.
\label{transition_radius}
\end{equation}
Which means, even if the core velocity \(v_0\) is smaller than the transition velocity \(v_\mathrm{t}\), in the spike region it can increase, and can become greater than \(v_\mathrm{t}\). If that happens, we call that radius (\(r\)) the transition radius (\(r_\mathrm{t}\)). At \(r=r_\mathrm{t}\), the DM velocity becomes equal to the transition velocity, and the interaction changes from \(a=0\) to \(a=4\).

Thus, if the velocity in the core is less than the transition velocity (\(v_0<v_\mathrm{t}\)), moving inwards, from higher \(r\) to lower \(r\), we have the shallower spike with the exponent \(\gamma = 3/4\). Once we cross the transition radius moving inwards, we have the steeper spike with the exponent \(\gamma = 7/4\).

However, if the velocity in the core is greater than the transition velocity (\(v_0>v_\mathrm{t}\)), then inside the spike, moving inwards, the velocity will only increase. So, there won't be any transition. In this case, as \(v_0\) is always greater than the transition velocity, the only profile that we will have is the \(a=4\) profile.

Therefore, following the procedure from  ACD2024, we start by calculating the core parameters (\(v_0, r_1\)) for \(a=0\) case by solving Eq.~\eqref{poisson}, the mass constraint, and Eq.~\eqref{core_radius}. In Eq.~\eqref{core_radius}, for the quantity \(\langle \sigma v \rangle\), we use Eq.~(3) of  ACD2024 for \(a=0\) case
\begin{equation}
    \langle \sigma v \rangle = \sigma_0 v_0.
\end{equation}
We see that the cross section we used is independent of the velocity, in agreement with Eq.~\eqref{sigma_v_dep}.

If the core velocity calculated above turns out to be less than \(v_\mathrm{t}\), we can continue calculating the spike parameters (\(r_\mathrm{sp}, \gamma\)) using Eq.~\eqref{spike_radius} and Eq.~\eqref{gamma}. We can then also calculate the transition radius \(r_\mathrm{t}\) using Eq.~\eqref{transition_radius}.

However, if the velocity in the core is greater than the transition velocity (\(v_0>v_\mathrm{t}\)), then we have to recalculate the core parameters, this time using 
\begin{equation}
    \sigma = \sigma_0 \left(\frac{v_\mathrm{t}}{v_0}\right)^4.
\end{equation}
We used Eq.~(B5) from  ACD2024 for this expression. This means that
\begin{equation}
    \begin{aligned}
        \langle \sigma v \rangle &= \sigma_0 \left(\frac{v_\mathrm{t}}{v_0} \right)^4 \times v_0  \\
         &= \sigma_0 \frac{v_\mathrm{t}^4}{v_0^3}.
    \end{aligned}
\end{equation}
After recalculating core parameters (\(v_0, r_1\)) using this \(\langle \sigma v \rangle\), we calculate the spike parameters (\(r_\mathrm{sp}, \gamma\)) using Eq.~\eqref{spike_radius} and Eq.~\eqref{gamma} (this time for \(a=4\)).

Note that 
\begin{equation}
     \langle \sigma v \rangle = \left\{ 
    \begin{array}{ll}
        \sigma_0v_0, & v_0 < v_t \\
        \sigma_0 \frac{v_t^4}{v_0^3}. & v_0 > v_t\, .
    \end{array}
    \right.
\label{sigma_v_boht_cases}
\end{equation}

\begin{figure}[htb]
  \centering
  \includegraphics[width=0.45\textwidth]{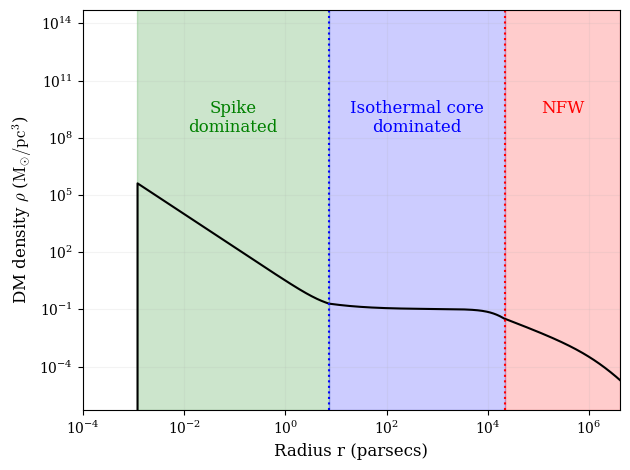}
  \caption{Density profile in three regions. We set the transition velocity \(v_\mathrm{t}=1000\) km/s, \((\sigma_0/m)\times(t_\mathrm{age}/1\,\mathrm{Gyr}) = 0.3\) \(\mathrm{cm^2/g}\), and (\(M, q, z) = (6\times10^9\mathrm{M}_\odot, 1, 0)\) where $M$ is the sum of the two black hole masses, $q$ is the ratio of the two black hole masses, and $z$  is the redshift.  We get the SIDM velocity dispersion in core \(v_0=497.2\) km/s, core radius \(r_1=2.2\times10^4\) pc, and transition radius \(r_\mathrm{t}=7.3\) pc. The cutoff radius, which is twice the Schwarzschild radius, \(r_\mathrm{cutoff}=4GM_\mathrm{tot}/c^2= 1.15\times10^{-3}\) pc. The transition from the NFW (red) to the isothermal core-dominated (blue) region happens at \(r=r_1\). The next transition to the density spike-dominated (green) region takes place at \(r=r_\mathrm{t}\).}
  \label{density_profile}
\end{figure}

We show an example density profile in Fig.~\ref{density_profile}. To maintain the continuity of the density profiles which undergo transition at \(r_\mathrm{t}\), we used \(r_\mathrm{t}\) as the scaling radius (\(r_\mathrm{sp}\)) in Eq.~\eqref{spike}. Also, to smoothen out the curves in the isothermal core dominated and the spike dominated regions, we plot \(\rho(r) = \rho_\mathrm{core}(r) + \rho_\mathrm{spike}(r)\), which smoothens the profile at the transition radius \(r_t\), and doesn't change it in the other regions.

\section{\label{sec:dynamics}Dynamics of a binary merger}
For the gravitational waves, we consider a supermassive black hole binary (SMBHB) merging in a circular orbit. The black holes of masses \(M_1\) and \(M_2\) are at a separation \(a\). The total energy of the system is given by the sum of gravitational potential energy and kinetic energy. Using the Newtonian approximation, we get
\begin{equation}
    E_\mathrm{tot} = -\frac{G M_1 M_2}{a} + \frac{1}{2} M_1 v_{1 , 2}^2 \,\,\,\,,
\label{tot_E}
\end{equation}
where \(v_{1, 2}\) is the relative velocity of the black hole of mass \(M_1\) from the frame of reference of the second one. The centripetal force is provided by the gravitational force.
\begin{equation}
    \begin{aligned}
        \frac{M_1 v_{1, 2}^2}{a} &= \frac{G M_1 M_2}{a^2} \\
         v_{1, 2}^2 &= \frac{G M_2}{a}.
    \end{aligned}
\end{equation}

Substituting \(v_{1, 2}^2\) in Eq.~\eqref{tot_E}, we get,
\begin{equation}
    \begin{aligned}
        E_\mathrm{tot} &= -\frac{G M_1 M_2}{a} + \frac{1}{2} M_1 \frac{G M_2}{a} \\
         &= -\frac{G M_1 M_2}{2a}.
    \end{aligned}
\end{equation}
The separation \(a\) decreases with time due to the binary inspiral. Differentiating the equation above with respect to time, we get
\begin{equation}
    \begin{aligned}
        \dot{E}_\mathrm{tot} \equiv P_\mathrm{tot} &= -\frac{G M_1 M_2}{2}(-1)a^{-2} \dot{a} \\
        &= \frac{G M_1 M_2}{2a^2}\dot{a},
    \label{ptot_m1m2}
    \end{aligned}
\end{equation}
where \(\dot{a}\) is \(\mathrm{d}a/\mathrm{d}t\). In our analysis, we consider two ways of energy loss for this system. Gravitational wave radiation and the dynamical friction caused by the binary moving in the SIDM halo.
\subsection{Gravitational wave energy loss}
For the energy loss due to gravitational wave radiation, we follow \cite{peters1964}. In \cite{peters1964}, Eq.~(5.4) gives us
\begin{equation}
    \frac{\mathrm{d}E}{\mathrm{d}t} \equiv P_\mathrm{gw} = - \frac{32}{5} \frac{G^4 M_1^2 M_2^2 (M_1 + M_2)}{c^5 a^5}
\label{pgw_m1m2}
\end{equation}

\subsection{Dynamical friction energy loss}
We use Eq.~(4) in  ACD2024, which uses Chandrasekhar's dynamical frictional force to obtain the power lost in dynamical friction as
\begin{multline}
    P_\mathrm{df} = 12 \pi q^2 \sqrt{1 + q} (G M_1)^{3/2} \sqrt{a} \\
    \times \left( \frac{N_1(q)}{q^3} \rho(r_1) + N_2(q) \rho(r_2) \right).
\label{pdf_m1m2}
\end{multline}
The Coulomb logarithm that appears in the dynamical friction expression is set to 3 following  ACD2024. Here, \(M_1\) is the mass of the heavier black hole, the primary SMBH. Then, \(q\) is the mass ratio of the secondary to the primary SMBH, \(q=M_2/M_1 \), which is \(\le 1\). The distances of the primary and secondary black holes from the origin (centre of mass) are \(r_1\) and \(r_2\), respectively. The functions \(N_1(q)\) and \(N_2(q)\) are given in Eq.~(C2) and Eq.~(C3) of  ACD2024 as
\begin{equation}
    N_i = \mathrm{erf} \left( \frac{u_i}{\sqrt{2}} \right) - \sqrt{\frac{2}{\pi}} \, u_i \, e^{-u_i^2 / 2},
\end{equation}
and
\begin{equation}
    u_i = 
    \begin{cases}
    \frac{11}{4} \, q^{3/2} (1 + q)^{-3/2}, & i = 1 \\
    \frac{11}{4} \, (1 + q)^{-3/2}. & i = 2
    \end{cases}
\end{equation}
Thus, \(P_\mathrm{df}\) depends on the SIDM halo density at \(r_1\) and \(r_2\).

\subsection{Total energy loss}
In the rest of the analysis, we will use the variables total black hole mass \(M\) and the mass ratio \(q\) instead of \(M_1\) and \(M_2\). We rewrite Eq.~\eqref{ptot_m1m2}, Eq.~\eqref{pgw_m1m2}, and Eq.~\eqref{pdf_m1m2} using these variables. \(M_1 + M_2\) can simply be rewritten as \(M\). For \(M_1M_2\), we use \(M_2 = q M_1\) and get
\begin{equation}
    \begin{aligned}
        &M_1 + M_2 = M \\
        &M_1 +  q M_1 = M\\
        &M_1 (1+q) = M.
    \end{aligned}
\label{m1m2m}
\end{equation}
Hence, \(M_1=M/(1+q)\) and \(M_2=qM/(1+q)\), which results in \(M_1M_2 = qM^2/(1+q)^2\).

Similarly, for \(r_1\) and \(r_2\), utilising the fact that the origin is at the centre of mass, we can write
\begin{equation}
    \begin{aligned}
        & M_1 r_1 = M_2 r_2 \\
        & r_1 = \frac{M_2}{M_1} r_2 \\
        & r_1 = q r_2.
    \end{aligned}
\end{equation}
We also know that \(r_1 + r_2 =a\). Substituting for \(r_1\), we get
\begin{equation}
    \begin{aligned}
        &q r_2 + r_2 = a \\
        & (q+1) r_2 = a,
    \end{aligned}
\end{equation}
which gives us \(r_2 = a/(1+q)\) and \(r_1= qa/(1+q)\). Substituting this in Eq.~\eqref{ptot_m1m2}, we get
\begin{equation}
    \begin{aligned}
        P_\mathrm{tot} \equiv P_\mathrm{gw} + P_\mathrm{df} 
        = \frac{G }{2a^2}\frac{qM^2}{(1+q)^2}\dot{a},
    \end{aligned}
\end{equation}
which gives \(\dot{a}\) as
\begin{equation}
    \dot{a} = \left( P_\mathrm{gw} + P_\mathrm{df} \right) \frac{2 (1+q)^2}{GqM^2} a^2.
\label{dadt}
\end{equation}
This quantity is essential in the gravitational strain calculation.

\section{\label{sec:strain}Gravitational strain spectrum}
The gravitational strain of a single binary source is given in Eq.~(6) of  Agazie2023,
\begin{equation}
    h_\mathrm{s}^2(f)  = \frac{32}{5c^8} \frac{(G\mathcal{M})^{10/3}}{d_c^2} (2\pi f_p)^{4/3},
\end{equation}
where \(d_c\) is the comoving distance to the source as a function of redshift \(z\), \(\mathcal{M}\) is the binary chirp mass, \(f\) is the observer-frame gravitational wave frequency, and \(f_p\) is the source-frame orbital frequency of the binary inspiral. If the source is at redshift \(z\), these two frequencies are related by
\begin{equation}
    f = \frac{2f_p}{(1+z)}.
\end{equation}
Moreover, the binary chirp mass given by Eq.~(1) of  Agazie2023,
\begin{equation}
    \mathcal{M} = \frac{M q^{3/5}}{(1+q)^{6/5}}.
\end{equation}

We are interested in all binary sources in our universe that have emitted GWs, which have a frequency \(f\) in the observer's frame. The observer-frame frequency depends on the binary separation, redshift, total black hole mass, and mass ratio. Thus, there can be many sources with different combinations of (\(M, q, z\)) at various separations \(a\), which can give us the required observer-frame GW frequency \(f\). We add the GW strains of all those sources. To do that, we follow Eq.~(5) of  Agazie2023,
\begin{equation}
    h_c^2(f) = \int dM \, dq \, dz \frac{\partial^4 N}{\partial M \partial q \partial z \partial \ln f_p} h_s^2(f_p),
\label{hc}
\end{equation}
where \(h_c(f)\) is the characteristic strain of the GWB over a logarithmic frequency interval.
From Eq.~(7) of  Agazie2023, we can write
\begin{equation}
    \frac{\partial^4 N}{\partial M \partial q \partial z \partial \ln f_p} = \frac{\partial^3 \eta}{\partial M \partial q \partial z} \cdot \tau(f_p) \cdot 4 \pi c \, (1 + z) \, d_c^2
\label{number}
\end{equation}
where $c$ is the speed of light. The first term on the right-hand side is the differential volumetric number density of binaries, which depends on \((M, q, z)\). It gives the number of binaries of mass \(M\), mass ratio \(q\), and redshift \(z\) that would be present in a unit volume. The \(\tau(f_p)\) is described as the binary hardening timescale, which is the time (measured in the source frame, i.e. the center of mass rest-frame of the SMBHB) spent in a given logarithmic frequency interval. It can be written as
\begin{equation}
    \begin{aligned}
        \tau(f_p) &= \frac{\partial t}{\partial \mathrm{{ln}} f_p} \\
         &= \frac{\partial t}{\partial f_p}\frac{\partial f_p}{\partial \mathrm{{ln}} f_p} \\
       &= \frac{\partial t}{\partial f_p} f_p  \\
         &= \frac{f_p}{\dot{f_p}},
    \end{aligned}
\end{equation}
where \(\dot{f_p} = \partial f_p/ \partial t \).
\subsection{Binary hardening timescale}
To write the hardening timescale in terms of the separation \(a\) and \(\dot{a}\), we write the Newtonian force equation (from the frame of reference of \(M_2\)),
\begin{equation}
    \begin{aligned}
        &M_1 \omega^2 a = \frac{G M_1 M_2}{a^2}\\
        &\omega^2 = \frac{G M_2}{a^3},
    \end{aligned}
\end{equation}
where the orbital angular frequency is related to the orbital frequency as \(\omega = 2\pi f_p\). Substituting this, we get
\begin{equation}
    \begin{aligned}
        &\left(2\pi f_p \right)^2 = \frac{G M_2}{a^3} \\
        &f_p^2 = \frac{G M_2}{4\pi^2} a^{-3} \\
        &f_p = \left(\frac{G M_2}{4\pi^2}\right)^{1/2} a^{-3/2}.
    \end{aligned}
\end{equation}
Taking a time derivative, we get
\begin{equation}
    \begin{aligned}
        \dot{f_p}& = \left(\frac{G M_2}{4\pi^2}\right)^{1/2} \left( -\frac{3}{2}\right) a^{-5/2} \frac{\mathrm{d} a}{\mathrm{d} t} \\
         &= \left(\left(\frac{G M_2}{4\pi^2}\right)^{1/2} a^{-3/2}\right) \left( -\frac{3}{2}\right) a^{-1} \dot{a} \\
        & = f_p \left( -\frac{3}{2}\right) a^{-1} \dot{a},
    \end{aligned}
\end{equation}
which gives
\begin{equation}
    \tau(f_p) \equiv \frac{f_p}{\dot{f_p}}  = -\frac{3}{2}\frac{a}{\dot{a}}.
\end{equation}
Combining the above equation with Eq.~\eqref{hc} and \eqref{number}, the characteristic strain can be written as
\begin{equation}
    \begin{aligned}
        h_c^2(f) = \int dM \, dq \, dz  &\frac{\partial^3 \eta}{\partial M \partial q \partial z} \cdot \left(-\frac{3}{2}\frac{a}{\dot{a}}\right) \cdot 4 \pi c \, (1 + z) \, d_c^2 \\
        &\times \frac{32}{5c^8} \frac{(G\mathcal{M})^{10/3}}{d_c^2} (2\pi f_p)^{4/3}.
    \end{aligned}
\end{equation}

We already obtained \(\dot{a}\) for our model in Eq.~\eqref{dadt}. Using all of this, we can write the characteristic strain \(h_c(f)\) in terms of  \(M\),  \(q\), \(z\), and \(f\). We need to perform the integration in Eq.~\eqref{hc} numerically for each frequency. Following  Agazie2023, we consider the limits of integration to be \(M: (10^4 \mathrm{M}_\odot,10^{12}  \mathrm{M}_\odot)\), \(q: (0.001, 1)\), and \(z: (0.001, 10)\). The only remaining quantity to express in terms of (\(M, q, z\)) is \(\partial^3 \eta/\partial M \partial q \partial z\).

\subsection{Binary population}
Eq.~(22) in  Agazie2023 states that
\begin{equation}
    \frac{\partial^3 \eta}{\partial M\, \partial q\, \partial z} = \frac{\partial^3\eta_{\text{gal-gal}}}{\partial M_{\star 1}\, \partial q_{\star}\, \partial z} \frac{\partial M_{\star 1}}{\partial M} \frac{\partial q_{\star}}{\partial q},
\end{equation}
where \(M_{\star 1}\) is the stellar mass of the primary merging galaxy, and \(q_{\star}\) is the stellar mass ratio of the secondary to the primary galaxy. The relation between \(M_{\star 1}\) and \(M_1\) is given in Eq.~\eqref{mbh_mbulge} and Eq.~\eqref{mbulge_mstar}. Using that, we can obtain \(\partial M_{\star 1}/\partial M\). For a black hole of mass \(M_{\mathrm{BH}}\), the relation between the corresponding galaxy stellar mass \(M_{\star}\) is given by
\begin{equation}
    \begin{aligned}
        &\log_{10}\left(\frac{M_{\mathrm{BH}}}{\mathrm{M}_\odot}\right) = \mu + \alpha_\mu \log_{10} \left( \frac{0.615 \cdot M_{\star}}{10^{11}\, \mathrm{M}_\odot} \right) + \mathcal{N}(0, \epsilon_\mu).\\
    \end{aligned}
\end{equation}
Rearranging the terms, we get
\begin{equation}
    \begin{aligned}
        &\alpha_\mu\log_{10} M_{\star} = \log_{10}M_{\mathrm{BH}} - \log_{10} \mathrm{M}_\odot \\
        &\quad\quad\quad\quad\quad\quad- \mu - \mathcal{N}(0, \epsilon_\mu) - \alpha_\mu \log_{10}  \left( \frac{0.615}{10^{11}\, \mathrm{M}_\odot} \right).
    \end{aligned}
\label{mstar_mbh}
\end{equation}
Differentiating both sides with respect to \(M_\text{BH}\) we obtain
\begin{equation}
    \begin{aligned}
        &\frac{\partial}{\partial M_{\mathrm{BH}}}\left( \alpha_\mu \log_{10} M_{\star}\right) = \frac{\partial}{\partial M_{\mathrm{BH}}} \log_{10}M_{\mathrm{BH}}  \\
        &\alpha_\mu \frac{1}{M_{\star}} \frac{\partial M_{\star}}{\partial M_{\mathrm{BH}}} = \frac{1}{M_{\mathrm{BH}}} \\
        &\frac{\partial M_{\star}}{\partial M_{\mathrm{BH}}} = \frac{1}{\alpha_\mu} \frac{M_{\star}}{M_{\mathrm{BH}}}.
    \end{aligned}
\end{equation}
In this relation, if we replace \(M_{\mathrm{BH}}\) with the mass of our primary merging SMBH \(M_1\), and \(M_\star\) with the corresponding stellar mass of the primary galaxy \(M_{\star 1}\), we can write,
\begin{equation}
    \begin{aligned}
        &\frac{\partial M_{\star 1}}{\partial M_1} = \frac{1}{\alpha_\mu} \frac{M_{\star 1}}{M_1}
    \end{aligned}
\end{equation}
However, we want \(\partial M_{\star 1}/\partial M\). Hence, applying the chain rule, we obtain
\begin{equation}
    \begin{aligned}
        \frac{\partial M_{\star 1}}{\partial M} &=  \frac{\partial M_{\star 1}}{\partial M_1} \frac{\partial M_1}{\partial M} \\
    \end{aligned}
\end{equation}
And, we already know from Eq.~\eqref{m1m2m} that \(M_1 = M/(1+q)\). Differentiating, we get
\begin{equation}
    \frac{\partial M_1}{\partial M} = \frac{1}{(1+q)}.
\end{equation}
Therefore, 
\begin{equation}
    \begin{aligned}
        \frac{\partial M_{\star 1}}{\partial M} &= \frac{1}{\alpha_\mu} \frac{M_{\star 1}}{M_1} \: \frac{1}{(1+q)} \\
        \frac{\partial M_{\star 1}}{\partial M} &= \frac{1}{\alpha_\mu} \frac{M_{\star 1}}{M}.        
    \end{aligned}
\end{equation}
Now, for \(\partial q_{\star}/\partial q\), we perform a similar analysis. Using Eq.~\eqref{mstar_mbh} for \(M_{\star2}\) and \(M_{\star1}\), followed by a subtraction, we get
\begin{equation}
    \begin{aligned}
        \alpha_\mu\log_{10} M_{\star 2} - \alpha_\mu\log_{10} M_{\star 1} &= \log_{10}M_2 - \log_{10}M_1 \\
        \alpha_\mu \log_{10} \left( \frac{M_{\star 2}}{M_{\star 1}} \right) & = \log_{10} \left( \frac{M_2}{M_1} \right) \\
        \log_{10} \left( \frac{M_{\star 2}}{M_{\star 1}} \right)^{\alpha_\mu} & = \log_{10} \left( \frac{M_2}{M_1} \right) \\
        \left( \frac{M_{\star 2}}{M_{\star 1}} \right)^{\alpha_\mu} & = \frac{M_2}{M_1} \\
        q_{\star}^{\alpha_\mu} &= q.
    \end{aligned}
\end{equation}
Differentiating with respect to \(q\), we get
\begin{equation}
    \begin{aligned}
        \frac{\partial}{\partial q} q_{\star}^{\alpha_\mu} &= \frac{\partial q}{\partial q} \\
        \alpha_\mu q_{\star}^{\alpha_\mu - 1} \frac{\partial q_{\star}}{\partial q} &= 1 \\
        \frac{\partial q_{\star}}{\partial q} &= \frac{1}{\alpha_\mu} q_{\star}^{1 -\alpha_\mu}.
    \end{aligned}
\end{equation}
Next, for \(\partial^3\eta_{\text{gal-gal}}/\partial M_{\star 1}\, \partial q_{\star}\, \partial z\), we refer to Eq.~(13) from  Agazie2023. It states,
\begin{equation}
    \frac{\partial^3 \eta_{\text{gal-gal}}}{\partial M_{\star 1} \, \partial q_{\star} \, \partial z} =
    \frac{\Psi(M_{\star 1}, z')}{M_{\star 1} \ln(10)} \cdot
    \frac{P(M_{\star 1}, q_{\star}, z')}{T_{\text{gal-gal}}(M_{\star 1}, q_{\star}, z')}
    \cdot \frac{\partial t}{\partial z'}.
\label{d3eta}
\end{equation}
Here, \(z'\) is the redshift at the time before the merger starts. Everything on the RHS is calculated at this pre-merger redshift \(z'\). The redshift that we use on the LHS is the redshift after the merger is complete. i.e. \(z' = z'[t]\) and \(z = z[t + T_{\text{gal-gal}}]\). We have the redshift-time cosmological relation
\begin{equation}
    \frac{\mathrm{d}t}{\mathrm{d}z} = \frac{1}{(1 + z) H(z)}
\end{equation}
Using this, along with Eq.~\eqref{Hz}, the post-merger redshift can be found if the pre-merger redshift \(z\) and the galaxy merger time \(T_{\text{gal-gal}}\) are known. We do this by solving for \(z_\mathrm{post-merger}\).
\begin{equation}
    T_{\text{gal-gal}} = \int_{t}^{t + T_{\text{gal-gal}}} dt = 
    \int_{z_\mathrm{post-merger}}^{z_\mathrm{pre-merger}}(1 + z) H(z) \, dz
\end{equation}
For calculating \(\Psi(M_{\star 1}, z')\), \(P(M_{\star 1}, q_{\star}, z')\), and \(T_{\text{gal-gal}}(M_{\star 1}, q_{\star}, z')\), we use Eq.~(15), Eq.~(16), Eq.~(19), and Eq.~(20) from  Agazie2023:
\begin{equation}
    \Psi(M_{\star 1}, z') = \ln(10) \, \Psi_0 \left[ \frac{M_{\star 1}}{M_{\psi}} \right]^{\alpha_{\psi}} \exp\left( -\frac{M_{\star 1}}{M_{\psi}} \right)
\end{equation}
where
\begin{equation}
    \begin{aligned}
        &\log_{10} \left( \Psi_0 / \mathrm{Mpc}^{-3} \right) = \psi_0 + \psi_z \cdot z', \\
        &\log_{10} \left( M_{\psi} / M_{\odot} \right) = m_{\psi,0} + m_{\psi,z} \cdot z', \\
        &\alpha_{\psi} = 1 + \alpha_{\psi,0} + \alpha_{\psi,z} \cdot z'.
    \end{aligned}
\label{gsmf}
\end{equation}
The galaxy pair fraction is calculated using
\begin{equation}
    P(M_{\star 1}, q_{\star}, z') = P_0 \left( \frac{M_{\star 1}}{10^{11} M_{\odot}} \right)^{\alpha_p} (1+z')^{\beta_p} q_{\star}^{\gamma_p},
    \end{equation}
where
\begin{equation}
    \begin{aligned}
        &\alpha_p = \alpha_{p,0} + \alpha_{p,z} \cdot z', \\
        &\beta_p = \beta_{p,0} + \beta_{p,z} \cdot z', \\
        &\gamma_p = \gamma_{p,0} + \gamma_{p,z} \cdot z'.
    \end{aligned}
\end{equation}
For the galaxy merger time, we use
\begin{equation}
        T_{\mathrm{gal-gal}}(M_{*1}, \, q_{*}, \, z') = T_{0} \left( \frac{M_{*1}}{10^{11} \, M_{\odot}/h} \right)^{\alpha_{t}} (1 + z')^{\beta_{t}} \, q^{\gamma_{t}}
\end{equation}
where
\begin{equation}
    \begin{aligned}
        &\alpha_t = \alpha_{t,0} + \alpha_{t,z} \cdot z' \\
        &\beta_t = \beta_{t,0} + \beta_{t,z} \cdot z' \\
        &\gamma_t = \gamma_{t,0} + \gamma_{t,z} \cdot z'.
    \end{aligned}
\end{equation}
Here, \(h\) is the dimensionless Hubble parameter given by
\begin{equation}
    h = \frac{H_0}{100 \, \mathrm{km \, s^{-1} \, Mpc^{-1}}}.
\end{equation}

The values used for all these parameters are given in Table \ref{table_param_values}. Some of these parameters are varied in our analysis. The prior ranges for those parameters are also present in Table \ref{table_param_values}.

\begin{table}[htbp]
    \centering
    {
    \renewcommand{\arraystretch}{1.5}
    \begin{tabular}{|c|c|c|}
        \hline
        \textbf{Symbol} & \textbf{Fiducial Value} & \textbf{Priors} \\
        \hline
        $v_t$        & $\cdots$     & $\mathcal{U}(1, 2000)$ km/s \\
        $\frac{\sigma_0}{m}\frac{t_\mathrm{age}}{1\,\mathrm{Gyr}}$   & $\cdots$  & $\mathcal{U}(0.01, 200)$ $\mathrm{cm^2/g}$ \\
        \hline
        $\psi_0$        & $\cdots$     & $\mathcal{N}(-2.56, 0.4)$ \\
        $\psi_z$         & $-0.60$      & $\cdots$ \\
        $m_{\psi,0}$    & $\cdots$     & $\mathcal{N}(10.9, 0.4)$ \\
        $m_{\psi,z}$    & $+0.11$      & $\cdots$ \\
        $\alpha_{\psi,0}$ & $-1.21$     & $\cdots$ \\
        $\alpha_{\psi,z}$ & $-0.03$     & $\cdots$ \\
        \hline
        $P_0$           & $+0.033$     & $\cdots$ \\
        $\alpha_{P,0}$  & $0.0$        & $\cdots$ \\
        $\alpha_{P,z}$  & $0.0$        & $\cdots$ \\
        $\beta_{P,0}$   & $+1.0$       & $\cdots$ \\
        $\beta_{P,z}$   & $0.0$        & $\cdots$ \\
        $\gamma_{P,0}$  & $0.0$        & $\cdots$ \\
        $\gamma_{P,z}$  & $0.0$        & $\cdots$ \\
        \hline
        $T_0$           & $+0.5$ Gyr   & $\cdots$ \\
        $\alpha_{T,0}$  & $0.0$        & $\cdots$ \\
        $\alpha_{T,z}$  & $0.5$        & $\cdots$ \\
        $\beta_{T,0}$   & $-0.5$       & $\cdots$ \\
        $\beta_{T,z}$   & $0.0$        & $\cdots$ \\
        $\gamma_{T,0}$  & $-1.0$       & $\cdots$ \\
        $\gamma_{T,z}$  & $0.0$        & $\cdots$ \\
        \hline
        $\mu$           & $\cdots$     & $\mathcal{N}(8.6, 0.2)$ \\
        $\alpha_\mu$    & $+1.10$      & $\cdots$ \\
        $\epsilon_\mu$  & $\cdots$     & $\mathcal{N}(0.32, 0.15)$ dex \\
        $f_{\star, \mathrm{bulge}}$ & $+0.615$ & $\cdots$ \\
        \hline
    \end{tabular}
    }
\caption{Priors and fiducial values for the model parameters}
\label{table_param_values}
\end{table}

With this, we now have everything needed to calculate the characteristic strain spectrum \(h_c(f)\).

\section{\label{sec:stats}Statistical analysis}
Our statistical analysis aims to find the posterior distribution of the parameters of our model using the GWB data from the PTA. The dataset we used was the NANOGrav-15 year dataset.

\subsection{Parameters of our model}
We vary 6 parameters in our analysis corresponding to different parts of our model. For the SIDM profile, \(v_t\) and \((\sigma_0/m)\times(t_\mathrm{age}/1\,\mathrm{Gyr})\) are varied. Transition velocity, $v_t$, dictates where the density profile should transition to a steeper spike. The density of the spike affects the power lost due to dynamical friction. Thus, $v_t$ controls how much power is lost in dynamical friction. And, \((\sigma_0/m)\times(t_\mathrm{age}/1\,\mathrm{Gyr})\) appears in Eq.~\eqref{core_radius}. 

The differential number density of galaxy mergers is expressed in terms of galaxy stellar mass function (GSMF) \(\Psi(M_{\star}, z')\) in Eq.~\eqref{d3eta}. Following  Agazie2023, we vary \(\psi_0\) and \(m_{\psi, 0}\) from Eq.~\eqref{gsmf} which are used in describing the GSMF.

In  Agazie2023, two parameters from the black hole mass-bulge mass relation are varied. Following that, we vary \(\mu\) and \(\epsilon_{\mu}\) from Eq.~\eqref{mbh_mbulge}.

The priors on these parameters are listed in Table \ref{table_param_values}.

\subsection{Library generation and MCMC}
We use \texttt{holodeck} (described in  Agazie2023) to perform library generation and MCMC chain creation. 8000 unique combinations of the 6 model parameters are generated following their prior distributions and ranges using Latin hypercube sampling. For each of these 8000 combinations, a corresponding strain spectrum is calculated. This is called the library. An example of how a library looks is shown in Table \ref{table_library}.

\begin{table}[htbp]
    \centering
    {
    \renewcommand{\arraystretch}{1.5}
    \begin{tabular}{|c|c c c c c c|c|}
        \hline
         & \(v_t\) & \(\frac{\sigma_0}{m}\frac{t_\mathrm{age}}{1\,\mathrm{Gyr}}\) & \(\psi_0\) & \(m_{\psi, 0}\) & \(\mu\) & \(\epsilon_{\mu}\) & \(h_c(f)\) \\
        \hline
        1 & 500 & 10 & -2.4 & 10.5 & 8.6 & 0.3 & [\(2 \times 10^{-14}\), ..., \(3 \times 10^{-15}\)] \\
        2 & 600 & 5 & -2.3 & 11.1 & 8.8 & 0.4 & [\(8 \times 10^{-15}\), ..., \(6 \times 10^{-16}\)] \\
        \vdots & \vdots & \vdots & \vdots & \vdots & \vdots & \vdots & \vdots \\
        \vdots & \vdots & \vdots & \vdots & \vdots & \vdots & \vdots & \vdots \\
        8000 & 100 & 0.1 & -2.5 & 10.8 & 8.7 & 0.2 & [\(4 \times 10^{-14}\), ..., \(9 \times 10^{-15}\)] \\
        \hline
    \end{tabular}
    }
\caption{Example: The structure of the library}
\label{table_library}
\end{table}

Once the library generation is done, a Gaussian process (GP) is used to interpolate between the points in the library. Depending on how big the parameter space is, one might need to adjust how many points should be in the library to train the GP. This enables us to rapidly generate MCMC chains for the parameters of the model. The Gaussian process ensures that for each new point sampled in MCMC, one doesn't need to calculate the strain numerically. The strain can simply be interpolated using the trained GP.

For the choice of dataset, we followed  Agazie2023. We used the NANOGrav 15-year dataset with HD (Hellings Downs) correlated free spectrum modelled simultaneously with additional MP (monopole-correlated), DP (dipole-correlated) red noise, and CURN (common uncorrelated red-noise). We used the strain values from the five lowest frequency bins of this dataset. These five frequency bins have the best constraints on the strain derived from PTA data. These strain values (or rather distributions, as there is a Gaussian distribution in the strain derived from the dataset in each frequency bin) are used in calculating likelihoods, posterior distributions, MCMC chains, etc.

\subsection{Multiple realizations}
Another important detail of our analysis is multiple realizations of the number of binaries. For numerically performing the integration in the strain calculation from Eq.~\eqref{hc}, we divide the mass, mass ratio, and the redshift ranges into a finite number of bins (90, 80, 100, respectively, for example). Using Eq.~\eqref{number}, we can get the differential number in each bin. When it is multiplied by that bin's \(dM \, dq \, dz\), we obtain the number of binaries in that bin. Let's say we obtained a number for a specific bin (\(M, q, z\)). Following  Agazie2023, we generate multiple realizations of this number by drawing from its Poisson distribution. In our analysis, we created 2000 realizations for a number in each bin.

This gives us the number of binaries in that (\(M, q, z\)) bin, which produce GWs of observer-frame frequency \(f\). If this number is multiplied by the strain from a single source corresponding to that specific (\(M, q, z\)) bin, we get the total strain contribution from that (\(M, q, z\)) bin. Thus, in our library, we have 8000 parameter combinations. For each of those combinations, strain values are calculated for the 5 lowest PTA frequencies as mentioned before. As there are 2000 realizations, the shape of our library is (8000, 5, 2000).

\texttt{holodeck} helps facilitate these steps of library generation, Gaussian process interpolation, and MCMC chain generation.

\section{\label{sec:results}Results}

In our analysis, we probed the parameters of the SIDM model which solves the ``final parsec problem". In  ACD2024, they gave constraints on the parameters \(v_t\) and \((\sigma_0/m)\times(t_\mathrm{age}/1\,\mathrm{Gyr})\) such that the ``final parsec problem" can be solved. These constraints ensured that the SIDM spike wouldn't be disrupted by the energy transferred from the SMBHB. We implemented these constraints in our analysis, too.

We used the MCMC method to analyse the parameters of our SIDM model. The MCMC sample points follow the posterior distribution, which is obtained by calculating the prior distribution and likelihood. The likelihood is calculated by comparing the characteristic strain spectra from the NANOGrav 15-year data with the strain for the parameters in the parameter space. In our analysis, we used \(\sim\)60,000 MCMC samples.

\begin{figure*}[htb]
  \centering
  \includegraphics[width=0.9\textwidth]{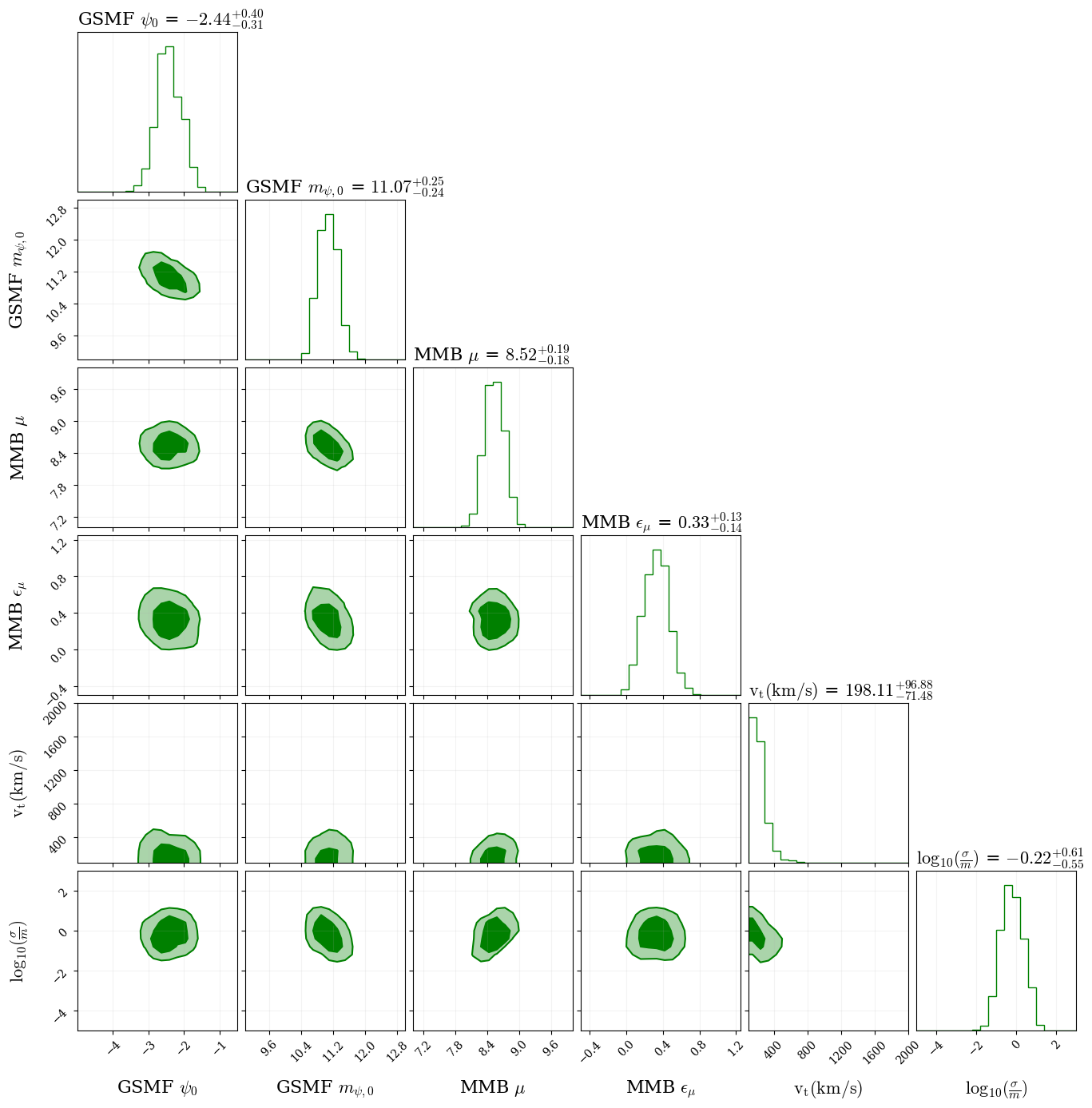}
  \caption{Posterior distribution of the model parameters with 68\% and 95\% confidence interval contours. Reported values correspond to the medians and their 68\% credible intervals. \(\sigma/m\) is calculated for \((M, q, z) = (3.61\times10^{9}\mathrm{M}_\odot, 0.96, 0.46)\).}
  \label{posterior}
\end{figure*}

The posterior distribution for the six parameters of the model is presented in Fig.~\ref{posterior}. The first five parameters were varied directly in our analysis in addition to the parameter \((\sigma_0/m)\times(t_\mathrm{age}/1\,\mathrm{Gyr})\). These posterior distributions are plotted after applying the ``final-parsec problem solution constraints". The 95\% region for  \(\psi_0\) matches well with the results of  ACD2024 who found a best fit of $\psi_0 \sim-2.5$.   Our results for (\(\psi_0\), \(m_{\psi, 0}\), \(\mu\), \(\epsilon_{\mu}\)) are in agreement with the results from  Agazie2023's Fig.~10 for their phenomenological model with astrophysical priors. We have used the same priors in our analysis. 

We obtained \(\sigma/m\) from \((\sigma_0/m)\times(t_\mathrm{age}/1\,\mathrm{Gyr})\) by using Eq.~\eqref{sigma_v_boht_cases} that gives us \(\sigma_0\) as
\begin{equation}
     \sigma = \left\{ 
    \begin{array}{ll}
        \sigma_0, & v_0 < v_t \\
        \sigma_0 \left(\frac{v_t}{v_0}\right)^4. & v_0 > v_t
    \end{array}
    \right.
\end{equation}
We calculate \(t_\mathrm{age}\) using the method described in  ACD2024, where they calculated \(t_\mathrm{age}\) by setting it equal to the dynamical friction hardening timescale \(t_\mathrm{df}\). In  ACD2024 \(t_\mathrm{df}\) is calculated by computing the inspiral duration for the binary to reach from a separation of 10 parsecs to 0.1 parsecs. The details of the calculation we followed are given in Appendix B of  ACD2024. The total mass, mass ratio, and the redshift values used to calculate \(t_\mathrm{age}\) and \(\sigma/m\) are \((M, q, z) = (3.61\times10^{9}\mathrm{M}_\odot, 0.96, 0.46)\). We chose these values because SMBHBs corresponding to these values make the most significant contribution to the GWB signal for the SIDM model in the lowest PTA frequency band. We determined the value corresponding to the most significant contribution by subtracting the integrand of Eq.~\eqref{hc} for the highest posterior SIDM model minus the integrand for the GW-only model (Agazie2023) and determined which value of \((M, q, z)\) gave the greatest absolute difference. 

To implement the final-parsec problem solution constraint, we excluded the grey region in Fig.~5 in  ACD2024. For numerical implementation, we use the conditions \((\sigma_0/m)\times(t_\mathrm{age}/1\,\mathrm{Gyr}) > 0.25\) and \(t_\mathrm{df} > 50\)~Myr. 
\textcolor{black}{
We assume other astrophysical effects bring the SMBHB to a \(\sim\) 1 parsec separation. So we start our simulations from a 1 parsec separation, and the dynamical friction by the SIDM then goes on to drive the binary inspiral to smaller separations.}

\begin{figure}[ht]
  \centering
  \includegraphics[width=0.45\textwidth]{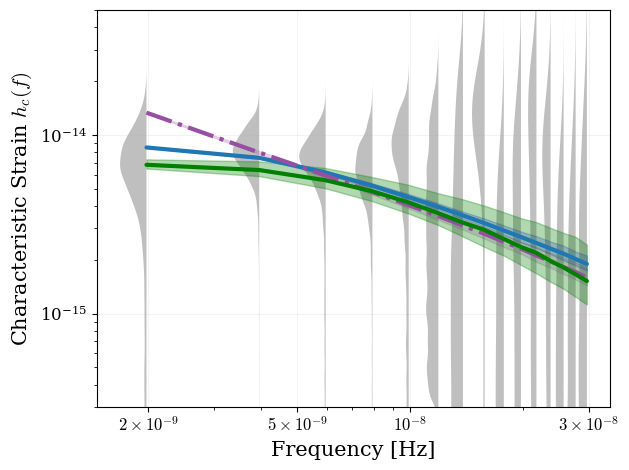}
  \caption{\textbf{Green}: Characteristic strain calculated for the maximum posterior parameters. We considered 2000 realizations. Median values are plotted with the solid line with a 68\% confidence interval. \textbf{Grey}: NANOGrav 15-year data with HD (Hellings Downs) correlated free spectrum modelled simultaneously with additional MP (monopole-correlated), DP (dipole-correlated) red noise, and CURN (common uncorrelated red-noise), also denoted as HD-w/MP+DP+CURN in \cite{gwb_nanograv, Agazie_2023}. \textbf{Blue}: Best fit strain spectra for the Phenomenological model from  Agazie2023. \textbf{Purple}: Best fit strain spectra for the GW-only model from the Fig. 1 (right panel) of  Agazie2023 }
  \label{strain_maxL}
\end{figure}

We present the characteristic strain calculated for the parameters with the maximum posterior value in Fig.~\ref{strain_maxL}, alongside the strain from the NANOGrav 15-year GWB data. The maximum posterior value parameters, which were used to calculate the strain spectra, are (\(\psi_0\), \(m_{\psi, 0}\), \(\mu\), \(\epsilon_{\mu}\), \(v_t\), \((\sigma_0/m)\times(t_\mathrm{age}/1\,\mathrm{Gyr})\)) = (\textcolor{black}{-2.45, 11.22, 8.47, 0.32, 133.34 km/s, 1.68 \(\mathrm{cm^2/g}\)}). \cg{In that figure, we also show the GW-only model and the phenomenological model, which has two additional environmental parameters for a target binary lifetime and a small-separation hardening index.}

\begin{figure}[ht]
  \centering
  \includegraphics[width=0.45\textwidth]{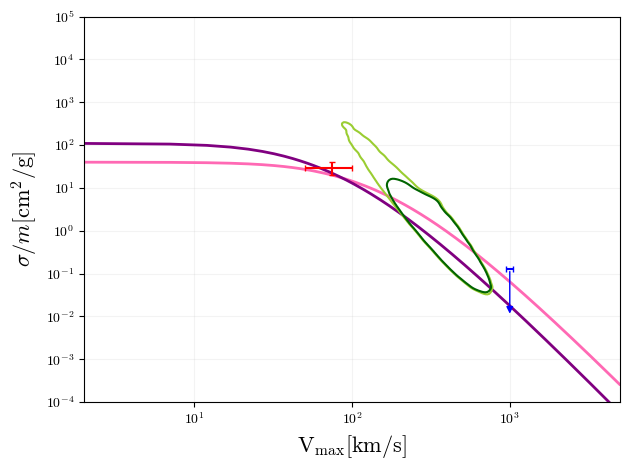}
  \includegraphics[width=0.45\textwidth]{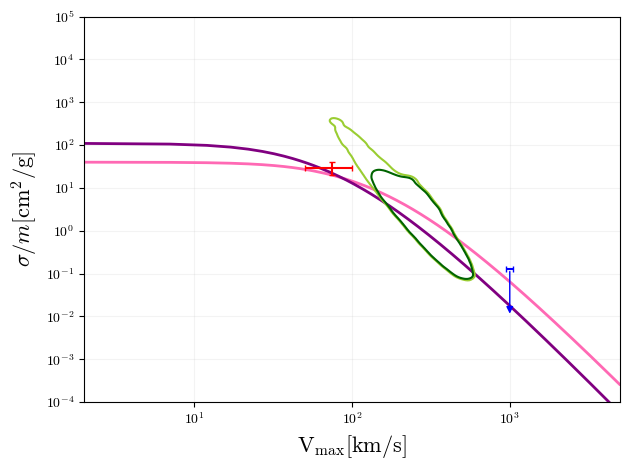}
  \includegraphics[width=0.45\textwidth]{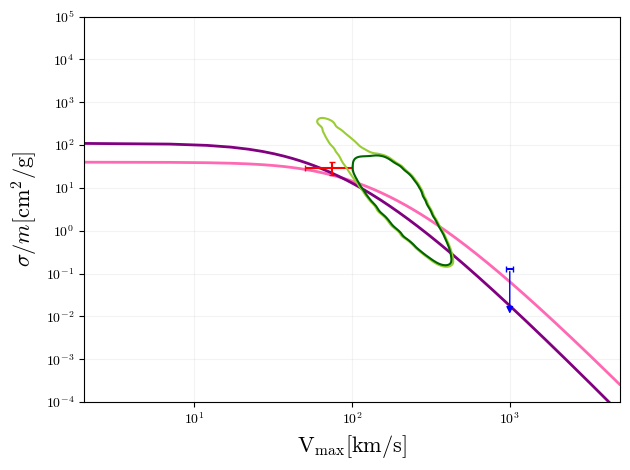}
  \caption{Comparison with \(\sigma/m\) from \cite{roberts2025}: \(\sigma/m\) is calculated for \((M,q, z) = (2.40 \times 10^9\mathrm{M}_\odot, 0.96 , 0.96)\), \((1.59 \times 10^9\mathrm{M}_\odot, 0.96, 0.87)\), and \((8.62 \times 10^8\mathrm{M}_\odot, 0.96, 1.05)\) for the top, middle and bottom panels respectively. Results from this work are shown in \textbf{green}. The light‑green region is the 95\% region obtained by considering the constraints derived from GWB PTA data. The dark green region corresponds to incorporating constraints given by the ``final parsec problem solution" as well as the GWB PTA constraints. \textbf{Pink and purple}: Results from \cite{roberts2025} Fig. 9 right panel. Pink and purple lines correspond to a dark matter mass of 0.3 GeV and 0.5 GeV, respectively, which are consistent with galaxy rotation curve and cluster strong lensing data. The \(V_\mathrm{max}\) quantity on the x-axis is related to the core velocity \(v_0\) as \(V_\mathrm{max} = v_0 / 0.64\). \textbf{Red and blue}: Points from \cite{roberts2025} Fig. 9 right panel. Red point corresponds to the values derived in \cite{roberts2025} for the dwarf galaxies (\(V_\mathrm{max} = 75\pm25\)km/s, \(\sigma/m = 30\pm10\) \(\mathrm{cm^2/g}\)), and the blue point corresponds to the values for galaxy groups and cluster scales (\(V_\mathrm{max} = 1000\pm55\)km/s, \(\sigma/m < 0.13\) \(\mathrm{cm^2/g}\)).}
  \label{roberts2025_n8000_1e10}
\end{figure}

In Fig.~9 (right panel) of \cite{roberts2025}  \(\sigma/m\) is plotted against the velocity, \(\mathrm{V_{max}}\), which relates to velocity dispersion as \(v_0=0.64 \,\mathrm{V_{max}}\). We calculated \(v_0\) in Section \ref{subsec:core} by solving Eq.~\eqref{poisson}, Eq.~\eqref{core_radius}, and the mass constraint. We follow the same method as \cite{roberts2025}, and plot \(\sigma/m\) versus \(\mathrm{V_{max}}\) using our MCMC chains. In Fig.~\ref{roberts2025_n8000_1e10}, we plot a 95\% confidence level contour and lay it over the \cite{roberts2025} results for  several different choices of $M$.
As can be seen, even though the 95\% region moves for different masses, it is still consistent with the results from \cite{roberts2025}.

We found the total masses, mass ratios, and redshifts of SMBHBs which gave the most significant contribution for the SIDM model in the three of the five lowest frequency bins of the GWB signal corresponding to the maximum posterior parameter values. The corresponding \(\sigma/m\) values and error bars obtained for these different SMBHBs are given in Table~\ref{tab:sidm_constraints}.
As can be seen \(\sigma/m\) is well constrained from below, however, the GWB data and the ``final parsec problem" solution condition don't lead to a good upper constraint.

\begin{table}[htbp]
\renewcommand{\arraystretch}{1.4}
\centering
\caption{Median 68\% credible‐interval constraints on the self-interaction cross–section per unit mass \(\sigma/m\) and the maximum circular velocity \(V_{\max}\) for three most contributing SMBHB total masses $M$, mass ratios $q$, and the redshift $z$.}
\begin{tabular}{ccc|cc}
\hline\hline
\(\mathrm{M \,[M_\odot]}\) & \(q\) & \(z\) & \(\sigma/m\;[\mathrm{cm^{2}\,g^{-1}}]\) & \(V_{\max}\;[\mathrm{km\,s^{-1}}]\)\\
\hline
\(2.40\times10^{9}\) & 96 & 0.96 & \(0.84^{+5.85}_{-0.74}\)  & \(195^{+104}_{-76}\)\\
\(1.59\times10^{9}\) & 0.96 & 0.87 & \(1.64^{+10.37}_{-1.44}\)  & \(187^{+99}_{-78}\)\\
\(8.62\times10^{8}\) & 0.96 & 1.05 & \(3.99^{+21.55}_{-3.53}\) & \(180^{+98}_{-78}\)\\
\hline
\end{tabular}
\label{tab:sidm_constraints}
\end{table}

However, as we can see in Fig.~\ref{roberts2025_n8000_1e10}, after applying the ``final-parsec problem solution" constraints, the allowed region reduces compared to the only ``PTA constraints" region. Interestingly, for the ``final-parsec" solution, higher values of \(\sigma/m\) are not allowed. This is expected because higher  \(\sigma/m\) causes larger cores, which results in less steep spikes, leading to smaller dynamical friction. Thus, higher \(\sigma/m\) causes lower dynamical friction. Therefore, for the final-parsec problem to be solved, \(\sigma/m\) is restricted to lower values.

\section{\label{sec:conclusion}Conclusion}
We analysed an \cg{SIDM} model with a central density spike as it presents a solution to the ``final parsec problem". The presence of \cg{SIDM} around a merging supermassive black hole binary would alter the resulting gravitational wave emission. This effect can be observed, thanks to the recent \cg{GWB PTA} 
 data. We used this data to probe the parameters of the self-interacting dark matter model with a density spike.

We simulated the gravitational characteristic strain spectra for galaxy and supermassive black hole mergers. \cg{This enabled us to generate}  multiple strain spectra by varying the model parameters, and \cg{we} used that information to fit the observed data. \cg{Using MCMC methods}, with the condition of solving the ``final parsec problem", gave a posterior distribution of the model parameters. 
For a variety of different black hole binary total masses, we found \(\sigma/m\) and \(V_{\rm max}\) values that were consistent with recent galaxy rotation curve and cluster strong lensing constraints on self-interacting dark matter \cite{roberts2025}.

\cg{As seen in Fig.~\ref{strain_maxL}, including environmental effects can also produce effects similar to SIDM on the strain spectrum. It is difficult to distinguish between these effects solely from the GWB data. Future independent observations that probe SIDM, like galaxy rotation curves or strong lensing, have the potential to provide further evidence for the SIDM model \cite{roberts2025}. This would further strengthen the case that the final parsec problem is solved by SIDM.}

Complementary analyses that include eccentricity or alternative DM configurations find qualitatively similar departures from a pure GW‑driven spectrum \cite{Bi:2023tib,Ellis:2023dgf,Shen:2023pan,Hu:2023oiu}.
If the SMBHB is driven to high eccentricities, then even at a high separation, gravitational wave (GW) radiation can cause the binary to merge more quickly. Stellar interactions can cause this eccentricity. In \cite{chen_2024}, they consider three-body scattering and eccentric orbits. It would be interesting to evaluate how this mechanism was affected by the presence of SIDM.

In general, DM spikes are subject to astrophysical uncertainties from stellar and baryon effects~\cite{Ullio:2001fb}.
Also, in ref.~\cite{Jiang2023}, they argue that the DM halo response to baryons is more diverse in SIDM than in CDM and depends sensitively on galaxy size. In future work, we would like to incorporate baryon effects in our evaluation of the impact of SIDM on PTA data. 


\cgg{Recent work has also explored the prospects for identifying SMBHB candidates as a high-energy neutrino source class using IceCube data, and discussed the potential connection to nano-Hz GW observations with PTAs \cite{Jaroschewski_2022, Pugazhendh2025}. 
 In our context, an SIDM spike changes the binary’s residence time in the PTA band and therefore can affect the overlap between PTA-loud systems and any neutrino-bright (gas/jet-rich) sub-population.

Ref.~\cite{DosopoulouSilk2025} evaluated how an inspiralling BH binary inside a DM spike can perturb the spike and thereby modulate the gamma-ray flux from DM annihilation while emitting detectable GWs. It would be interesting to investigate how SIDM would change this signal.}

\section*{Acknowledgments}
We gratefully acknowledge support by the Marsden Fund Council grant MFP-UOA2131 from New Zealand Government funding, managed by the Royal Society Te Ap\={a}rangi. We would also like to thank Gonzalo Alonso-Alvarez, Marco Galoppo, Chris Harvey-Hawes, Morag Hills, Emma Johnson, Manoj Kaplinghat, Luke Kelley, Zachary Lane, Pierre Mourier, Michael Ryan, and El Mehdi Zahraoui for helpful discussions. We also acknowledge the University of Canterbury Research Cluster facilities for providing computational resources that significantly improved the efficiency of our computations (\href{https://doi.org/10.18124/CANTERBURYNZ-UCRCH}{DOI:10.18124/CANTERBURYNZ-UCRCH}, RRID:SCR\_027870).

\section*{Data availibility}
The code and the data used for producing the results of this paper are publicly avaliable \cite{tiruvaskar_2026_zenodo}.

\section*{Note added}
After this manuscript was posted to the \texttt{arXiv}, we became aware of the
independent study by M.-C.~Chen and Y.~Tang \cite{Chen:2025jch}.  Where our 
work overlaps with theirs, the qualitative conclusions are broadly consistent.

\clearpage %
\bibliography{apssamp}%

\end{document}